\documentclass[12pt]{article}%
\usepackage{ amsmath, graphics, rotating}%

\begin{document}
\large
\begin{center}{\large\bf SUPERSYMMETRY IN QUANTUM THEORY OVER 
A GALOIS FIELD}
\end{center}
\vskip 1em \begin{center} {\large Felix M. Lev} \end{center}
\vskip 1em \begin{center} {\it Artwork Conversion Software Inc.,
1201 Morningside Drive, Manhattan Beach, CA 90266, USA 
(E-mail:  felixlev@hotmail.com)} \end{center}
\vskip 1em

{\it Abstract:}
\vskip 0.5em

As shown in our previous papers (hep-th/0209001
and reference therein), quantum theory based on a 
Galois field (GFQT) possesses a new symmetry between 
particles and antiparticles, which has no analog in 
the standard approach. In the present paper it is 
shown that this symmetry (called the AB one) is 
also compatible with supersymmetry.
We believe this is a strong argument in favor of our
assumption that the AB symmetry is a fundamental
symmetry in the GFQT (and in nature if it is
described by quantum theory over a Galois field).
We also consider operatorial formulations of space
inversion and X inversion in the GFQT. It is shown
in particular that the well known fact, that
the parity of bosons is real and the parity of
fermions is imaginary, is a simple consequence
of the AB symmetry.

\begin{flushleft} PACS: 02.10.Lh, 03.65.Bz, 03.65.Ca, 11.30.Ly\end{flushleft}

\section{Introduction}
\label{S1} 
 
In papers \cite{lev1} we have proposed an approach to
quantum theory where the wave functions of the system
under consideration are described by elements of a linear 
space over a Galois field, and the operators of physical 
quantities - by linear operators in this space. A detailed
discussion of this approach has been given in a recent paper
\cite{lev2}. In particular, it has been shown that when
the characteristic $p$ of the Galois field is very large,
there exists a correspondence between such a description
and the standard one (in terms of states and operators in
the Hilbert space). In other words, there exist
situations when quantum theory based on a Galois field
(GFQT) predicts results which are practically
indistinguishable from those given by the standard theory. 
It has also been argued that the
description of quantum systems in terms of Galois fields 
is more natural than the standard description in terms of
complex numbers. 

The first obvious conclusion about the GFQT is as follows: 
since any Galois field has only a finite number of elements, 
divergencies in this theory cannot exist in 
principle and all operators are automatically well defined.
It is also natural to expect that, since arithmetic of Galois 
field differs from the standard one, the GFQT has some 
properties which have no analog in the standard theory.  

In particular, as shown in Ref. \cite{lev3}, the GFQT
possesses a new symmetry between particles and 
antiparticles called AB symmetry. The existence of this
symmetry is a consequence of the fact that in the GFQT
an irreducible representation (IR) of the symmetry
algebra describes a particle and its antiparticle
simultaneously. This automatically explains the
existence of antiparticles and shows that a particle and
its antiparticle are different states of the same object.
As argued in Refs. \cite{lev2,lev3}, such an 
explanation is more natural than the standard one, where
negative energy solutions of covariant equations are used
and a particle and its antiparticle are described by
independent IRs of the symmetry algebra. 

If the AB symmetry is combined with the Pauli 
spin-statistics theorem \cite{Pauli} then one comes
to the conclusions \cite{lev3} that in the GFQT
any neutral particle (in particular the photon) 
cannot be elementary but only composite, and
any interaction can contain only an even number
of creation/annihilation operators.

These results are in striking disagreement with our
experience but as argued in Ref. \cite{lev3} this 
by no means excludes a possibility that future
quantum physics will be based on a Galois field.

In the present paper a supersymmetric version
of the GFQT is considered. We assume that the symmetry 
algebra is the 
Galois field analog of the superalgebra osp(1,4), which is 
a supersymmetric generalization of the anti de Sitter (AdS) 
algebra so(2,3). Therefore our first goal is to construct
Galois field analogs of IRs of this superalgebra. This is
done in Sect. \ref{S4} by analogy with the construction of
Ref. \cite{Heidenreich} where all 
positive energy IRs of the osp(1,4) superalgebra in Hilbert
spaces have been found. In preparatory Sects. \ref{S2}
and \ref{S3}
we describe the most important properties of Galois
field analogs of IRs of the sp(2) and so(2,3) algebras.
In Sect. \ref{S5} explicit formulas for matrix elements
of fermionic operators of IR of the osp(1,4) algebra are
given. In Sect. \ref{S6} it is shown how the representation
operators in the space of IR can be related to the 
corresponding operators in the Fock space. The construction
is used in Sect. \ref{S7} where the main result of the
paper is proved: the AB symmetry is compatible with
supersymmetry. We believe this is a strong argument in favor
of our assumption (made in Ref. \cite{lev2}) that the
AB symmetry is a fundamental symmetry in the GFQT (and in
nature if it is described by quantum theory over a
Galois field). Finally, in Sects. \ref{S8} and \ref{S9}
we consider
operatorial formulation of space inversion and X inversion
in the GFQT. It is shown in particular that the well
known result, that the parity of bosons is real and the
parity of fermions is imaginary, is a direct consequence
of the AB symmetry.

For reading the present paper, only very elementary 
knowledge of Galois fields is needed. Although the
notion of the Galois field is extremely simple and
elegant, the majority of physicists is not familiar with
this notion. For this reason, in Ref. \cite{lev2} an
attempt has been made to explain the basic facts about
Galois fields in a simplest possible way (and using
arguments which, hopefully, can be accepted by physicists). 
The readers who are not familiar with Galois fields can 
also obtain basic knowledge from standard textbooks
(see e.g. Refs. \cite{VDW}). 
 
\section{Modular IRs of the sp(2) algebra}
\label{S2}

If a conventional quantum theory has a symmetry group or
algebra, then there exists a unitary representation of 
the group or a representation of the algebra by Hermitian
operators in the Hilbert space describing the quantum 
system under consideration. In the latter case the
representation is also often called unitary meaning
that it can be extended to the unitary representation
of the corresponding group. 

Let $p$ be a prime number and $F_{p^2}$ be a Galois field
containing $p^2$ elements. This field has only one
nontrivial automorphism $a\rightarrow {\bar a}$ (see e.g.
Refs. \cite{VDW,lev2}) which is the analog of complex
conjugation in the field of complex numbers. The automorphism
can be defined as $a\rightarrow {\bar a}= a^p$ \cite{VDW}. 

In the GFQT, unitary representations in Hilbert spaces are
replaced by representations in spaces over $F_{p^2}$. 
Representations in spaces over fields
of nonzero characteristics are called modular representations.
A review of the theory of modular IRs can be found e.g. in
Ref. \cite{FrPa}. In the present paper we do not need a
general theory since modular IRs in question can be
constructed explicitly.

A modular analog of the Hilbert
space is a linear space $V$ over $F_{p^2}$ supplied by a
scalar product (...,...) such that for any $x,y\in V$
and $a\in F_{p^2}$, $(x,y)\in F_{p^2}$
and the following properties are satisfied: 
\begin{equation}
(x,y) =\overline{(y,x)},\quad (ax,y)=\bar{a}(x,y),\quad 
(x,ay)=a(x,y)
\label{1}
\end{equation}   
By analogy with usual notations, we use $^*$ to denote 
the Hermitian conjugation in spaces over $F_{p^2}$. This means
that if $A$ is an operator in $V$ then $A^*$ is the operator
satisfying $(Ax,y)=(x,A^*y)$ for all $x,y\in V$.

By definition, a particle is called elementary if it is
described by an IR of the symmetry group or algebra in the
given theory. Since we assume that the role of the symmetry
algebra is played by the superalgebra osp(1,4), this implies 
that elementary particles are described by modular IRs of 
this superalgebra. For representations in Hilbert spaces,
all unitary positive energy IRs of the osp(1,4) superalgebra
have been found in Ref.
\cite{Heidenreich}. By analogy with this construction, we
will seek modular IRs of the osp(1,4) superalgebra by
investigating their decompositions into modular IRs of
the so(2,3) algebra.  In turn, the key role in 
constructing such modular IRs is played by modular IRs 
of the sp(2) subalgebra. The latter are also very 
important for understanding the
AB symmetry. On the one hand, modular IRs of the sp(2) 
algebra are very simple while on the other they clearly
demonstrate the main difference between the standard and
modular cases: in contrast to the standard case, where 
unitary IRs are necessarily infinite dimensional, all
modular IRs are finite dimensional (this statement has
been proved in the general case by Zassenhaus \cite{Zass}).

Representations of the sp(2) algebra are described by a
set of operators $(a',a",h)$ satisfying the 
commutation relations 
\begin{equation}
[h,a']=-2a'\quad [h,a"]=2a"\quad [a',a"]=h
\label{2}
\end{equation} 
The modular analogs of unitary representations of the
sp(2) algebra are characterized
by the conditions that $a^{'*}=a^"$ and $h^*=h$. 

The  Casimir operator of the second order for the algebra
(\ref{2}) has the form
\begin{equation}
K=h^2-2h-4a"a'=h^2+2h-4a'a"
\label{3}
\end{equation}

We will consider representations with the vector $e_0$, such that
\begin{equation}
a'e_0=0,\quad he_0=q_0e_0,\quad (e_0,e_0)=1 
\label{4}
\end{equation} 
One can easily prove \cite{lev1,lev2} that $q_0$ is
"real", i.e. $q_0\in F_p$ where $F_p$ is the residue field
modulo $p$: $F_p=Z/Zp$ where $Z$ is the ring of integers.
The field $F_p$ consists of $p$ elements and represents the
simplest possible Galois field.

Denote $e_n =(a")^ne_0$.
Then it follows from Eqs. (\ref{3}) and (\ref{4}),
that for any $n=0,1,2,...$
\begin{equation}
he_n=(q_0+2n)e_n,\quad Ke_n=q_0(q_0-2)e_n, 
\label{5}
\end{equation} 
\begin{equation}
a'a"e_n=(n+1)(q_0+n)e_n
\label{6}
\end{equation}
\begin{equation}
(e_{n+1},e_{n+1})=(n+1)(q_0+n)(e_n,e_n)
\label{7}
\end{equation}

The case $q_0=0$ is trivial and corresponds to zero representation,
so we assume that $q_0\neq 0$. Then we have the case when ordinary 
and modular representations considerably differ each other. 
Consider first the ordinary case when $q_0$ is any real positive 
number. Then IR is infinite-dimensional, $e_0$ is a vector
with a minimum eigenvalue of the operator $h$ (minimum weight)
and there are no vectors with the maximum weight. This is in 
agreement with the well known fact that unitary IRs of
noncompact groups are infinite dimensional. However in the
modular case $q_0$ is one of the numbers $1,...p-1$. 
The set $(e_0,e_1,...e_N)$ will be a basis
of IR if $a"e_i\neq 0$ for $i<N$ and $a"e_N=0$. These conditions
must be compatible with $a'a"e_N=0$. Therefore, as follows from
Eq. (\ref{6}), $N$ is defined by the condition $q_0+N=0$ in
$F_p$. As a result, $N=p-q_0$ and the dimension of IR is equal
to $p-q_0+1$.  

One might say that $e_0$ is the vector with the
minimum weight while $e_N$ is the vector with the maximum weight.
However, the notions of "less than" or "greater than" have only
a limited sense in $F_p$, as well as the notion of positive
and negative numbers in $F_p$. If $q_0$ is positive in this sense
(see Ref. \cite{lev2} for details), then Eqs. (\ref{6}) and
(\ref{7}) indicate that the modular IR under consideration can be 
treated as the modular analog of IR with "positive energies".
However, it is easy to see that $e_N$ is the
eigenvector of the operator $h$ with the eigenvalue $-q_0$ in
$F_p$, and the same IRs can be treated as the modular analog
of IRs with "negative energies" (see Ref. \cite{lev2} for
details). 

\section{Modular IRs of the so(2,3) algebra}
\label{S3}

The standard AdS group
is ten-parametric, as well as the Poincare group. However, 
in contrast to the Poincare group, all the representation 
generators are angular momenta. In Ref. \cite{lev2} we 
explained the reason why for our
purposes it is convenient to work with the units 
$\hbar/2=c=1$. Then the representation generators are
dimensionless, and the commutation relations for them can be 
written in the form
\begin{equation}
[M^{ab},M^{cd}]=-2i (g^{ac}M^{bd}+g^{bd}M^{cd}-
g^{ad}M^{bc}-g^{bc}M^{ad})
\label{8}
\end{equation}
where $a,b,c,d$ take the values 0,1,2,3,5, and the operators 
$M^{ab}$ are antisymmetric. The diagonal metric tensor has the 
components $g^{00}=g^{55}=-g^{11}=-g^{22}=-g^{33}=1$.  
In these units the spin of fermions is odd, and the spin of 
bosons is even. If $s$ is the particle spin then the 
corresponding IR of the su(2) algebra has the dimension
$s+1$. Note that if $s$ is interpreted in such a way then it
does not depend on the choice of units (in contrast to the
maximum eigenvalue of the $z$ projection of the spin
operator).    

For analyzing IRs implementing Eq. (\ref{8}), it is convenient to
work with another set of ten operators. Let $(a_j',a_j",h_j)$ 
$(j=1,2)$ be two independent sets of operators satisfying the 
commutation relations for the sp(2) algebra
\begin{equation}
[h_j,a_j']=-2a_j'\quad [h_j,a_j"]=2a_j"\quad [a_j',a_j"]=h_j
\label{9}
\end{equation}
The sets are independent in the sense that
for different $j$ they mutually commute with each other. 
We denote additional four operators as $b', b",L_+,L_-$.
The meaning of $L_+,L_-$ is as follows. The operators 
$L_3=h_1-h_2,L_+,L_-$ satisfy the commutation relations
of the su(2) algebra
\begin{equation}
[L_3,L_+]=2L_+\quad [L_3,L_-]=-2L_-\quad [L_+,L_-]=L_3
\label{10}
\end{equation}
while the other commutation relations are as follows
\begin{eqnarray}
&[a_1',b']=[a_2',b']=[a_1",b"]=[a_2",b"]=\nonumber\\
&[a_1',L_-]=[a_1",L_+]=[a_2',L_+]=[a_2",L_-]=0\nonumber\\
&[h_j,b']=-b'\quad [h_j,b"]=b"\quad 
[h_1,L_{\pm}]=\pm L_{\pm},\nonumber\\
&[h_2,L_{\pm}]=\mp L_{\pm}\quad [b',b"]=h_1+h_2\nonumber\\
&[b',L_-]=2a_1'\quad [b',L_+]=2a_2'\quad [b",L_-]=-2a_2"\nonumber\\
&[b",L_+]=-2a_1",\quad [a_1',b"]=[b',a_2"]=L_-\nonumber\\
&[a_2',b"]=[b',a_1"]=L_+,\quad [a_1',L_+]=[a_2',L_-]=b'\nonumber\\
&[a_2",L_+]=[a_1",L_-]=-b"
\label{11}
\end{eqnarray}  
At first glance these relations might seem to be rather 
chaotic but in fact they are very natural in the Weyl basis
of the so(2,3) algebra. 

The relation between the above sets of ten operators is as follows
\begin{eqnarray}
&M_{10}=i(a_1"-a_1'-a_2"+a_2')\quad M_{15}=a_2"+a_2'-a_1"-a_1'\nonumber\\
&M_{20}=a_1"+a_2"+a_1'+a_2'\quad M_{25}=i(a_1"+a_2"-a_1'-a_2')\nonumber\\
&M_{12}=L_3\quad M_{23}=L_++L_-\quad M_{31}=-i(L_+-L_-)\nonumber\\
&M_{05}=h_1+h_2\quad M_{35}=b'+b"\quad M_{30}=-i(b"-b')
\label{12}
\end{eqnarray}
In addition, if 
$L_+^*=L_-$, $a_j^{'*}=a_j"$, $b^{'*}=b"$ and $h_j^*=h_j$ then
the operators $M^{ab}$ are Hermitian (we do not discuss the
difference between selfadjoined and Hermitian operators).   
In modular IRs of the so(2,3) algebra, the commutation relations 
(\ref{9}-\ref{11}) are realized in spaces over $F_{p^2}$
and the Hermitian conjugation is understood as explained above.

There exists a vast literature on ordinary IRs of the so(2,3) 
algebra in Hilbert spaces. The representations relevant to 
elementary particles in the AdS space have 
been constructed 
for the first time in Refs. \cite{Fronsdal,Evans}, while 
modular representations of algebra (\ref{9}-\ref{11}) have 
been investigated for the first time by Braden \cite{Braden}. 
In Refs. \cite{lev1,lev2} we have reformulated his 
investigation in such a way that the correspondence between 
modular and ordinary IRs are straightforward. Our construction
is described below. 

We use the basis in which the operators 
$(h_j,K_j)$ $(j=1,2)$ are diagonal. Here $K_j$ is the 
Casimir operator (\ref{3}) for algebra $(a_j',a_j",h_j)$. 
For constructing IRs we need operators relating different
representations of the sp(2)$\times$sp(2) algebra.
By analogy with Refs. \cite{Evans,Braden}, one of the
possible choices is as follows (see the discussion in Refs.
\cite{lev2,lev3})
\begin{eqnarray}
&B^{++}=b"-a_1"L_-(h_1-1)^{-1}-a_2"L_+(h_2-1)^{-1}
+\nonumber\\
&a_1"a_2"b'[(h_1-1)(h_2-1)]^{-1} \quad B^{+-}=L_+-a_1"b'(h_1-1)^{-1}\nonumber\\
&B^{-+}=L_--a_2"b'(h_2-1)^{-1}\quad B^{--}=b'
\label{13}
\end{eqnarray}

We consider the action of these operators only on the 
space of "minimal" 
sp(2)$\times$sp(2) vectors, i.e. such vectors $x$ that 
$a_j'x=0$ for $j=1,2$, and $x$ is the eigenvector of the
operators $h_j$. It is easy to see that if $x$ is a minimal 
vector such that
$h_jx=\alpha_jx$ then $B^{++}x$ is the minimal eigenvector of the
operators $h_j$ with the eigenvalues $\alpha_j+1$, $B^{+-}x$ - 
with the eigenvalues $(\alpha_1+1,\alpha_2-1)$, 
$B^{-+}x$ - with the eigenvalues $(\alpha_1-1,\alpha_2+1)$, 
and $B^{--}x$ - with the eigenvalues $\alpha_j-1$.

By analogy with the construction of ordinary representations with
positive energy \cite{Fronsdal,Evans}, we require the existence
of the vector $e_0$ satisfying the conditions
\begin{eqnarray}
&a_j'e_0=b'e_0=L_+e_0=0\quad h_je_0=q_je_0\nonumber\\
&(e_0,e_0)\neq 0\quad (j=1,2)
\label{14}
\end{eqnarray}

It is well known that $M^{05}=h_1+h_2$ is the AdS analog of 
the energy operator, since $M^{05}/2R$ becomes the usual 
energy when the AdS group is contracted to the Poincare one 
(here $R$ is the radius of the AdS space
while the notion of contraction has been developed in Ref.
\cite{IW}). As follows from 
Eqs. (\ref{9}) and (\ref{11}), the operators $(a_1',a_2',b')$ 
reduce the AdS energy by two units. Therefore in the 
conventional theory
$e_0$ is the state with the minimum energy. In this theory 
the spin in our units is equal to the maximum value of the 
operator $L_3=h_1-h_2$ in the
"rest state". For these reasons we use $s$ to denote $q_1-q_2$
and $m$ to denote $q_1+q_2$. 

In the standard classification
\cite{Fronsdal, Evans}, the massive case is characterized by
the conditions $q_1\geq q_2,\,\,q_2>1$ while the massless
one --- by $q_1\geq q_2,\,\,q_2=1$ 
(see also Ref. \cite{lev2}). There also exist two exceptional
IRs discovered by Dirac \cite{DiracS} and called Dirac
singletons. They are characterized by the values of $(q_1,q_2)$
equal to (1/2,1/2) and (3/2,1/2), respectively, i.e. the 
values of the mass
and spin are equal to (1,0) and (2,1), respectively. 
The quantity
1/2 in $F_p$ is a very big number if $p$ is big 
(since 1/2=$(p-1)/2$ in $F_p$) but nevertheless, the modular
analog of the singleton IRs can be investigated easily \cite{lev1}
and we will not dwell on this. As
explained above, in the modular case the notion of 'greater
than' is not so straightforward. Nevertheless, for IRs
related to elementary particles it is possible to formulate
an analog of these conditions \cite{lev2}. 

As follows from the above remarks, the elements 
\begin{equation}
e_{nk}=(B^{++})^n(B^{-+})^ke_0
\label{15}
\end{equation}
represent the minimal sp(2)$\times$sp(2) vectors with the 
eigenvalues of the operators $h_1$ and $h_2$ equal to 
$Q_1(n,k)=q_1+n-k$ and $Q_2(n,k)=q_2+n+k$, respectively.

We use $a(n,k)$ and $b(n,k)$ to denote the following quantities: 
\begin{equation}
a(n,k)=\frac{(n+1)(m+n-2)(q_1+n)(q_2+n-1)}{(q_1+n-k-1)(q_2+n+k-1)}
\label{16}
\end{equation}
\begin{equation}
b(n,k)=[(k+1)(s-k)(q_2+k-1)]/(q_2+n+k-1)
\label{17}
\end{equation}
Then it can be shown by a direct calculation (see Ref. 
\cite{lev2} for details) that
\begin{equation}
B^{--}B^{++}e_{nk}=a(n,k)e_{n,k}\quad 
(e_{n+1,k},e_{n+1,k})=a(n,k)(e_{nk},e_{nk})
\label{18}
\end{equation}
\begin{equation}
B^{-+}B^{++}e_{nk}=b(n,k)e_{nk}\quad (e_{n,k+1},e_{n,k+1})=
b(n,k)(e_{nk},e_{nk})
\label{19}
\end{equation}

In the massive case, as follows from Eqs. (\ref{17}) and
(\ref{19}), $k$ can assume only the values $0,1,...s$, as well
as in the ordinary case. At the same time, it follows from Eqs.
(\ref{16}) and (\ref{18}), that, in contrast to the ordinary 
case where $n=0,1,...\infty$, in the modular one $n=0,1,...n_{max}$. 
where $n_{max}=p+2-m$. Hence the space of minimal vectors has the 
dimension $(s+1)(n_{max}+1)$,
and IR turns out to be finite-dimensional and even finite since
the field $F_{p^2}$ is finite. In the massless case, when $q_2=1$,
the above expressions contain ambiguities $0/0$. The problem of
their correct treatment in the modular case has been discussed in Refs.
\cite{lev2,lev3} by analogy with the consideration in Ref. \cite{Evans}.
The result is that the values of $k$ are in the same range, for $k=0$
and $k=s$, $n$ takes the values $0,1,...,p+1-s$ while for the values
of $k$ in the range $1\leq k \leq s-1$ (such values of $k$ exist if 
$s\geq 2$) $n$ can take only the value $n=0$.

In this paper we describe in detail calculations in the massive 
case. The derivation in the massless and singleton cases is 
much simpler and the corresponding results are reported in 
Sect. \ref{S7}.
   
The full basis of the representation space can be chosen in the
form 
\begin{equation}
e(n_1n_2nk)=(a_1")^{n_1}(a_2")^{n_2}e_{nk} 
\label{20}
\end{equation}
where, as follows from the
results of this and preceding sections, 
\begin{eqnarray}
&n_1=0,1,...N_1(n,k)\quad n_2=0,1,...N_2(n,k)\nonumber\\
&N_1(n,k)=p-q_1-n+k\quad N_2(n,k)=p-q_2-n-k
\label{21}
\end{eqnarray}

As follows from Eqs. (\ref{7}) and (\ref{20}), the quantity 
\begin{equation}
Norm(n_1n_2nk)=(e(n_1n_2nk),e(n_1n_2nk))
\label{22}
\end{equation}
can be represented as
\begin{equation}
Norm(n_1n_2nk)=F(n_1n_2nk)G(nk)
\label{23}
\end{equation}
where 
\begin{equation}
F(n_1n_2nk)= n_1!(Q_1(n,k)+n_1-1)!n_2!(Q_2(n,k)+n_2-1)!
\label{24}
\end{equation}
\begin{equation}
G(nk)=(e_{nk},e_{nk})/[(Q_1(n,k)-1)!(Q_2(n,k)-1)!]
\label{25}
\end{equation}
By using Eqs. (\ref{16}-\ref{19}) and the definitions of 
$Q_1(n,k)$ and
$Q_2(n,k)$, one can show by a direct calculation that
\begin{eqnarray}
&G(nk)=[n!k!s!(e_0,e_0)][(s-k)!(q_1+n-k-1)!\nonumber\\
&(q_2+n+k-1)]^{-1}\prod_{l=1}^{n} \{[(m+l-3)(q_1+l-1)(q_2+l-2)]\nonumber\\
&[(q_1+l-k-2))(q_2+l+k-2)]^{-1}\} 
\label{26}
\end{eqnarray}

In standard Poincare and AdS theories there also exist IRs with
negative energies. They are not used in the standard approach and 
instead, for describing antiparticles one is using negative energy 
solutions of the corresponding covariant equation.
THe negative energy IRs can be constructed by analogy with 
positive energy ones.
Instead of Eq. (\ref{14}) one can require the existence of the
vector $e_0'$ such that
\begin{eqnarray}
&a_j"e_0'=b"e_0'=L_-e_0'=0\quad h_je_0'=-q_je_0'\nonumber\\
&(e_0',e_0')\neq 0\quad (j=1,2)
\label{27}
\end{eqnarray}
where the quantities $q_1,q_2$ are the same as for positive
energy IRs. It is obvious that positive and negative energy
IRs are fully independent since the spectrum of the operator
$M^{05}$ for such IRs is positive and negative, respectively.
At the same time, as shown in Refs. \cite{lev2,lev3},
{\it the modular analog of a positive energy IR 
characterized by $q_1,q_2$ in Eq. (\ref{14}), and the modular 
analog of a negative energy IR characterized by the same 
values of $q_1,q_2$ in Eq. (\ref{27}) represent the same
modular IR.} 

\section{Modular IRs of the osp(1,4) superalgebra}
\label{S4}

Representations of the osp(1,4) superalgebra have several interesting
distinctions from representations of the Poincare superalgebra. For
this reason we first briefly mention some well known facts about the
latter representations (see e.g Ref. \cite{Wein-super} for details).

Representations of the Poincare superalgebra are described by 14 
generators. Ten of them are the well known representation generators
of the Poincare group --- four momentum operators and six 
representation generators of the Lorentz group, which satisfy 
the well known commutation relations. In addition, there also 
exist four fermionic generators. The 
anticommutators of the fermionic generators are linear 
combinations of the momentum operators, and the commutators of 
the fermionic generators 
with the Lorentz group generators are linear combinations of the 
fermionic generators. In addition, the fermionic generators 
commute with the momentum operators. 

From the formal point of view, representations of the osp(1,4) 
superalgebra are also described by 14 generators --- ten 
representation generators of the so(2,3) algebra and four 
fermionic operators. There are three types of 
relations: the generators
of the so(2,3) algebra commute with each other as usual 
(see the preceding section), anticommutators of the 
fermionic generators are linear combinations of the so(2,3) 
generators and commutators of the latter with
the fermionic generators are their linear combinations. 
However, in fact representations of the osp(1,4) 
superalgebra can be described exclusively
in terms of the fermionic generators. The matter is 
as follows. In the general case the anticommutators of four 
operators form ten independent linear combinations. 
Therefore, ten bosonic generators can be expressed
in terms of fermionic ones. This is not the case for the 
Poincare superalgebra since the Poincare group generators 
are obtained from the so(2,3) ones by contraction. One can 
say that the representations of the
osp(1,4) superalgebra is an implementation of the idea that 
the supersymmetry
is the extraction of the square root from the usual 
symmetry (by
analogy with the well known treatment of the Dirac equation 
as a square
root from the Klein-Gordon one).

We denote the fermionic generators of the osp(1,4) superalgebra
as $(d_1,d_2,d_1^*,d_2^*)$ where the $^*$ means the 
Hermitian conjugation as 
usual. They should satisfy the following relations. 
If $(A,B,C)$ are any
fermionic generators, [...,...] is used to denote 
a commutator and
$\{...,...\}$ to denote an anticommutator then   
\begin{equation}
[A,\{ B,C\} ]=F(A,B)C + F(A,C)B
\label{30}
\end{equation}
where the form $F(A,B)$ is skew symmetric, $F(d_j,d_j^*)=1$ 
$(j=1,2)$ and
the other independent values of $F(A,B)$ are equal to zero.
The fact that the representation of the osp(1,4) superalgebra 
is fully defined by Eq. (\ref{30}) and the properties of the form
$F(.,.)$, shows that osp(1,4) is a special case of the
superalgebra.  

We can now {\bf define} the so(2,3) generators as follows:
\begin{eqnarray}
&b'=\{d_1,d_2\}\quad b"=\{d_1^*,d_2^*\}\quad
L_+=\{d_2,d_1^*\}\quad L_-=\{d_1,d_2^*\}\nonumber\\
&a_j'=(d_j)^2\quad a_j"=(d_j^*)^2\quad 
h_j=\{d_j,d_j^*\} \quad (j=1,2)
\label{31}
\end{eqnarray}
Then by using Eq. (\ref{30}) and the properties of the form
$F(.,.)$, one can show by a direct calculations that so 
defined operators satisfy the commutation relations
(\ref{9}-\ref{11}). This result can be treated as a fact
that the representation generators of the so(2,3) algebra
are not fundamental, only the fermionic generators are.

By analogy with the construction of IRs of the osp(1,4)
superalgebra in the conventional theory \cite{Heidenreich},
we require the existence of the cyclic vector $e_0$ satisfying the
conditions (compare with Eq. (\ref{14})):
\begin{eqnarray}
d_je_0=L_+e_0=0 \quad h_je_0=q_je_0\quad (e_0,e_0)\neq 0\quad (j=1,2)
\label{32}
\end{eqnarray}
The full representation space can be obtained by successively
acting by the fermionic generators on $e_0$ and taking all
possible linear combinations of such vectors.

We use $E$ to denote an arbitrary linear combination of the 
vectors $(e_0,d_1^*e_0,d_2^*e_0,d_2^*d_1^*e_0)$. Our next goal
is to prove a statement analogous to that in Ref. 
\cite{Heidenreich}: 

{\it Statement 1}: Any vector from the representation
space can be represented as a linear combination of the
elements $O_1O_2...O_nE$ where $n=0,1,,,$ and $O_i$ is a
generator of the so(2,3) algebra.

The first step is to prove a simple

{\it Lemma:} If $D$ is any fermionic generator then DE is a 
linear combination of elements $E$ and $OE$ where $O$ is a
generator of the so(2,3) algebra.

The proof is by a straightforward check using Eqs. 
(\ref{30}-\ref{32}). For example,
$$d_1^*(d_2^*d_1^*e_0)=\{d_1^*,d_2^*\}d_1^*e_0-d_2^*a_1"e_0=
b"d_1^*e_0-a_1"d_2^*e_0\,\, etc.$$

To prove Statement 1 we define the height of
a linear combination of the elements $O_1O_2...O_nE$ as the
maximum sum of powers of the fermionic generators in this
element. For example, since every generator of the so(2,3)
algebra is composed of two fermionic generators, the height
of the element $O_1O_2...O_nE$ is equal to $2n+2$ if $E$
contains $d_2^*d_1^*e_0$, is equal to $2n+1$ if $E$ does 
not contain $d_2^*d_1^*e_0$ but contains either $d_1^*e_0$ 
or $d_2^*e_0$ and is equal to $2n$ if $E$ contains only $e_0$.

We can now prove Statement 1 by induction. The elements
with the heights 0,1 and 2 obviously have the required form
since, as follows from Eq. (\ref{31}),
$d_1^*d_2^*e_0=b"e_0-d_2^*d_1^*e_0$. Let us assume that
Statement 1
is correct for all elements with the heights $\leq N$. 
Every element with the height $N+1$ can be represented as
$Dx$ where $x$ is an element with the height $N$. 
If $x=O_1O_2...O_nE$ then by using 
Eq. (\ref{30}) we can represent $Dx$ as $Dx=O_1O_2...O_nDE+y$
where the height of the element $y$ is equal to $N-1$. 
As follows from the induction assumption, $y$ has the required
form, and, as follows from Lemma, $DE$ is a linear
combination of the elements $E$ and $OE$. Therefore 
Statement 1 is proved.

As follows from Eqs. (\ref{30}) and (\ref{31}),
\begin{eqnarray}
&[d_j,h_j]=d_j\quad [d_j^*,h_j]=-d_j^*\quad (j=1,2)\nonumber\\
&[d_j,h_l]=[d_j^*,h_l]=0\quad (j,l=1,2\,\, j\neq l)
\label{33}
\end{eqnarray}
It follows from these expressions that if $x$ is such that
$h_jx=\alpha_jx$ $(j=1,2)$ then $d_1^*x$ is the eigenvector of the
operators $h_j$ with the eigenvalues $(\alpha_1+1,\alpha_2)$, 
$d_2^*x$ - with the eigenvalues $(\alpha_1,\alpha_2+1)$, 
$d_1x$ - with the eigenvalues $(\alpha_1-1,\alpha_2)$, 
and $d_2x$ - with the eigenvalues $\alpha_1,\alpha_2-1$.

Let us assume that $q_2\geq 1$ and $q_1\geq q_2$ in the
sense explained in the preceding section. We again use $m$
to denote $q_1+q_2$ and $s$ to denote $q_1-q_2$. 

Statement 1 obviously remains valid if we now assume that $E$
contains linear combinations of $(e_0,e_1,e_2,e_3)$ where 
\begin{eqnarray}
&e_1=d_1^*e_0\quad e_2=d_2^*e_0-\frac{1}{s+1}L_-e_1\nonumber\\ 
&e_3=(d_2^*d_1^*e_0-\frac{q_1-1}{m-2}b"+
\frac{1}{m-2}a_1"L_-)e_0
\label{34}
\end{eqnarray}
We assume for simplicity that $(e_0,e_0)=1$. Then
it can be shown by direct calculations using Eqs. 
(\ref{30}-\ref{32}) that 
\begin{equation}
(e_1,e_1)=q_1 \quad (e_2,e_2)=\frac{s(q_2-1)}{s+1}
\quad (e_3,e_3)=\frac{q_1(q_2-1)(m-1)}{m-2}
\label{35}
\end{equation}

As follows from Eqs. (\ref{30}-\ref{33}),  
$e_0$ satisfies Eq. (\ref{14}) and $e_1$ satisfies
the same condition with $q_1$ replaced by $q_1+1$. 
We see that the representation of the osp(1,4) superalgebra 
defined by Eq. (\ref{32}) necessarily contains at least two
IRs of the so(2,3) algebra characterized by the values of the
mass and spin $(m,s)$ and $(m+1,s+1)$, and the cyclic vectors
$e_0$ and $e_1$, respectively.

As follows from Eqs. (\ref{30}-\ref{33}), the vectors 
$e_2$ and $e_3$ satisfy the conditions 
\begin{eqnarray}
&h_1e_2=q_1e_2\quad h_2e_2=(q_2+1)e_2 \quad
h_1e_3=(q_1+1)e_3 \nonumber\\
&h_2e_3=(q_2+1)e_3\quad a_1'e_j=a_2'e_j=b'e_j=L_+e_j=0
\label{36}
\end{eqnarray}
$(j=2,3)$ and therefore (see Eq. (\ref{14})) they are 
candidates for
being cyclic vectors of IRs of the so(2,3) algebra if their
norm is not equal to zero. As follows from Eq. (\ref{35}),
$(e_2,e_2)\neq 0$ if $s\neq 0$ and $q_2\neq 1$. Therefore,
if these conditions are satisfied, $e_2$ is the cyclic vector
of IR of the so(2,3) algebra characterized by the values of
the mass and spin $(m+1,s-1)$. Analogously, if $q_2\neq 1$ then
$e_3$ is the cyclic vector of IR of the so(2,3) algebra 
characterized by the values of the mass and spin $(m+2,s)$. 

As already mentioned, our considerations are similar to those
in Ref. \cite{Heidenreich}. Therefore modular IRs of the 
osp(1,4) can be characterized in the same way as 
conventional IRs \cite{Heidenreich}:

\begin{itemize}

\item If $q_2>1$ and $s\neq 0$ (massive IRs), the osp(1,4)
supermultiplets contain four IRs of the so(2,3) algebra
characterized by the values of the mass and spin 
$$(m,s),(m+1,s+1),(m+1,s-1),(m+2,s).$$

\item If $q_2>1$ and $s=0$ (collapsed massive IRs), the osp(1,4)
supermultiplets contain three IRs of the so(2,3) algebra
characterized by the values of the mass and spin 
$$(m,s),(m+1,s+1),(m+2,s).$$

\item If $q_2=1$ (massless IRs) the osp(1,4)
supermultiplets contains two IRs of the so(2,3) algebra
characterized by the values of the mass and spin 
$$(2+s,s),(3+s,s+1)$$

\item Dirac supermultiplet containing two Dirac
singletons (see the preceding section).   
\end{itemize}

The first three cases have well known analogs of IRs of the 
super-Poincare algebra (see e.g. Ref. \cite{Wein-super}) while
there is no super-Poincare analog of the Dirac supermultiplet.

In this paper we describe details of calculation only for the
first case while the results for other cases are described in 
Sect. \ref{S7}.

We will say that the massive supermultiplet consists of particle 0,
particle 1, particle 2 and particle 3, where particle $i$ $(i=0,1,2,3)$
is described by the IR of the so(2,3) algebra with the cyclic vector
$e_i$. The set of minimal sp(2)$\times$sp(2) vectors in the multiplet
will be characterized by the elements $e(nk;i)$ where the range of $(n,k)$
for each $i$ is defined by the values of the corresponding mass and spin
$(m_i,s_i)$. As follows from the above classification, 
\begin{eqnarray}
&m_0=m\quad s_0=s\quad m_1=m+1\quad s_1=s+1\quad m_2=m+1\nonumber\\
&s_2=s-1\quad m_3=m+2\quad s_3=s
\label{37}
\end{eqnarray}
The full basis of the supermultiplet consists of 
elements $e(n_1n_2nk;i)$ 
where the range of $(n_1n_2nk)$ for each $i$ can be 
defined as described
in the preceding section. We also use $(q_{1i},q_{2i})$ to denote the
eigenvalues of the vectors $e_i$ with respect to $h_1$ and $h_2$, i.e.
$h_je_i=q_{ji}e_i$ $(j=1,2)$. It is easy to see that 
\begin{equation}
q_{10}=q_{12}=q_1\,\, q_{11}=q_{13}=q_1+1 \,\, q_{20}=q_{21}=q_2
\,\, q_{22}=q_{23}=q_2+1
\label{38}
\end{equation}
 
We now define
\begin{equation}
Norm(n_1n_2nk;i)=(e(n_1n_2nk;i),e(n_1n_2nk;i))\quad (i=0,1,2,3)
\label{39}
\end{equation}
Then one can show by analogy with Eqs. (\ref{24}-\ref{26}) that
\begin{eqnarray}
&Norm(n_1n_2nk;i)=F(n_1n_2nk;i)G(nk;i)\quad where\nonumber\\
&F(n_1n_2nk;i)= n_1!(q_{1i}+n-k+n_1-1)!n_2!(q_{2i}+n+k+n_2-1)!\nonumber\\
&G(nk;i)=[n!k!s_i!(e_i,e_i)][(s_i-k)!(q_{1i}+n-k-1)!\nonumber\\
&(q_{2i}+n+k-1)]^{-1}\prod_{l=1}^{n} \{[(m_i+l-3)(q_{1i}+l-1)(q_{2i}+l-2)]\nonumber\\
&[(q_{1i}+l-k-2))(q_{2i}+l+k-2)]^{-1}\} 
\label{40}
\end{eqnarray}

\section{Matrix elements of fermionic operators}
\label{S5}

As already noted, the main goal of the present 
paper is to prove that the representation operators
of the osp(1,4) superalgebra are compatible with the
AB symmetry. Since the representation operators of the
so(2,3) algebra are bilinear in fermionic operators
(see Eq. (\ref{31})), it is sufficient to prove the
compatibility for the latter. Moreover, it is 
sufficient to prove the compatibility only for the
operators $d_1^*$ and $d_2^*$ since $d_1$ and $d_2$
are adjoint to $d_1^*$ and $d_2^*$, respectively.   

By analogy with the way of finding matrix elements
in the so(2,3) case (see Sect. \ref{S3}), we can 
first investigate operators acting in the
subspace of minimal sp(2)$\times$sp(2) vectors.
Let us define
\begin{equation}
B_j^-=d_j\quad B_j^+=d_j^*-a_j"d_j(h_j-1)^{-1}\quad (j=1,2)
\label{41}
\end{equation}
By using Eqs. (\ref{30}) and (\ref{31}), it is
easy to show that if $x$ is a minimal 
vector such that $h_jx=\alpha_jx$ then 
$B_1^+x$ is the minimal eigenvector of the
operators $(h_1,h_2)$ with the eigenvalues 
$(\alpha_1+1, \alpha_2)$, 
$B_2^+x$ - with the eigenvalues 
$(\alpha_1,\alpha_2+1)$, 
$B_1^-x$ - with the eigenvalues 
$(\alpha_1-1,\alpha_2)$, and $B_2^-x$ - with the 
eigenvalues $(\alpha_1,\alpha_2-1)$. By using also 
Eq. (\ref{13}), one can show by a direct calculation that
\begin{eqnarray}
&[B_j^+,B^{++}]=[B_1^+,B^{+-}]=[B_2^+,B^{-+}]=
[B_j^-,B^{--}]=\nonumber\\
&[B_1^-,B^{-+}]=[B_2^-,B^{+-}]=0\quad (j=1,2)
\label{42}
\end{eqnarray}
In addition, if $x$ and $y$ are any minimal sp(2)$\times$sp(2)
vectors then
\begin{equation}
(x,B_j^+y)=(B_j^-x,y)\quad (j=1,2)
\label{43}
\end{equation}
This is a consequence of the definition (\ref{41}) and the fact
that $(a_j")^*x=a_j'x=0$.

By using these expressions, one can proceed as follows. First one
can compute $B_j^+e(0k;i)$ and $B_j^-e(0k;i)$ by using Eqs.
(\ref{13}), (\ref{15}), (\ref{30}), (\ref{31}) and the fact that,
as follows from Eq. (\ref{14}), $e(0k;i)=(L_-)^ke(00;i)$.
Then, by using Eq. (\ref{42}) and (\ref{43}), one can compute 
$B_j^+e(nk;i)$ and $B_j^-e(nk;i)$. When these calculations 
are done, one can compute $d_j^*e(n_1n_2nk;i)$ by 
using Eqs. (\ref{20}), (\ref{41}) and
the fact that the $d_j^*$ commute with $a_j"$:
\begin{eqnarray} 
&d_j^*e(n_1n_2nk;i)=(a_1")^{n_1}(a_2")^{n_2}d_j^*e(nk;i)=\nonumber\\
&(a_1")^{n_1}(a_2")^{n_2}[B_j^++a_j"B_j^-(h_j-1)^{-1}]
e(nk;i)\quad (j=1,2)
\label{44}
\end{eqnarray}

The final result is as follows.
\begin{eqnarray}
&d_1^*e(n_1n_2nk;0)=\frac{s+1-k}{s+1}e(n_1n_2nk;1)-\nonumber\\
&ke(n_1n_2,n,k-1;2)+\nonumber\\
&\frac{n(q_2+n-2)}{(q_1+n-k-1)
(q_1+n-k-2)(s+1)}e(n_1+1,n_2,n-1,k+1;1)+\nonumber\\
&\frac{n(q_1+n-1)}{(q_1+n-k-1)(q_1+n-k-2)}e(n_1+1,n_2,n-1,k;2)
\label{45}
\end{eqnarray}
\begin{eqnarray}
&d_1^*e(n_1n_2nk;1)=\frac{k(q_2+k-2)}{m-2}
e(n_1n_2,n+1,k-1;0)-\nonumber\\
&ke(n_1n_2n,k-1;3)+\nonumber\\
&\frac{(m+n-2)(q_1+n)(q_1-k-1)}{(m-2)(q_1+n-k)(q_1+n-k-1)}
e(n_1+1,n_2nk;0)+\nonumber\\
&\frac{n(q_1+n)}{(q_1+n-k)(q_1+n-k-1)}e(n_1+1,n_2,n-1,k;3)
\label{46}
\end{eqnarray}
\begin{eqnarray}
&d_1^*e(n_1n_2nk;2)=\frac{(s-k)(q_2+k+1)}{(s+1)(m-2)}
e(n_1n_2,n+1,k;0)-\nonumber\\
&\frac{s-k}{s+1}e(n_1n_2nk;3)-\nonumber\\
&\frac{(q_2+n-1)(m+n-2)(q_1-k-2)}{(q_1+n-k-1)(q_1+n-k-2)((s+1)(m-2)}
e(n_1+1,n_2,n,k+1;0)-\nonumber\\
&\frac{n(q_2+n-1))}{(q_1+n-k-1)(q_1+n-k-2)(s+1)}e(n_1+1,n_2,n-1,k+1;3)
\label{47}
\end{eqnarray}
\begin{eqnarray}
&d_1^*e(n_1n_2nk;3)=\frac{(q_2+k-1)(s+1-k)}{(m-2)(s+1)}
e(n_1n_2,n+1,k;1)-\nonumber\\
&\frac{k(q_2+k-1)}{m-2}e(n_1n_2,n+1,k-1;2)-\nonumber\\
&\frac{(m+n-1)(q_1-k-1)(q_2+n-1)}{(m-2)(q_1+n-k)(q_1+n-k-1)(s+1)}
e(n_1+1,n_2,n,k+1;1)-\nonumber\\
&\frac{(m+n-1)(q_1-k-1)(q_1+n)}{(m-2)(q_1+n-k)(q_1+n-k-1)}
e(n_1+1,n_2,n,k;2)
\label{48}
\end{eqnarray}
\begin{eqnarray}
&d_2^*e(n_1n_2nk;0)=e(n_1n_2nk;2)+\frac{1}{s+1}
e(n_1n_2n,k+1;1)+\nonumber\\
&\frac{n(s+1-k)(q_2+n-2)}{(s+1)(q_2+n+k-1)(q_2+n+k-2)}
e(n_1,n_2+1,n-1,k;1)-\nonumber\\
&\frac{kn(q_1+n-1)}{(q_2+n+k-1)(q_2+n+k-2)}e(n_1,n_2+1,n-1,k-1;2)
\label{49}
\end{eqnarray}
\begin{eqnarray}
&d_2^*e(n_1n_2nk;1)=\frac{q_1-k-1}{m-2}e(n_1n_2,n+1,k;0)+
e(n_1n_2nk;3)+\nonumber\\
&\frac{k(q_1+n)(m+n-2)(q_2+k-2)}{(m-2)(q_2+n+k-1)(q_2+n+k-2)}
e(n_1,n_2+1,n,k-1;0)-\nonumber\\
&\frac{kn(q_1+n)}{(q_2+n+k-1)(q_2+n+k-2)}e(n_1,n_2+1,n-1,k-1;3)
\label{50}
\end{eqnarray}
\begin{eqnarray}
&d_2^*e(n_1n_2nk;2)=-\frac{q_1-k-2}{(s+1)(m-2)}
e(n_1,n_2,n+1,k+1;0)-\nonumber\\
&-\frac{1}{s+1}e(n_1n_2n,k+1;3)+\nonumber\\
&\frac{(s-k)(m+n-2)(q_2+n-1)(q_2+k-1)}{(s+1)(m-2)(q_2+n+k)(q_2+n+k-1)}
e(n_1,n_2+1,nk;0)-\nonumber\\
&\frac{n(s-k)(q_2+n-1)}{(s+1)(q_2+n+k)(q_2+n+k-1)}e(n_1,n_2+1,n-1,k;3)
\label{51}
\end{eqnarray}
\begin{eqnarray}
&d_2^*e(n_1n_2nk;3)=-\frac{q_1-k-1}{m-2}
e(n_1,n_2,n+1,k;2)-\nonumber\\
&\frac{q_1-k-1}{(m-2)(s+1)}e(n_1n_2,n+1,k+1;1)+\nonumber\\
&\frac{(m+n-1)(q_2+k-1)}{(m-2)(q_2+n+k-1)(q_2+n+k)}
[\frac{(q_2+n-1)(s+1-k)}{s+1}e(n_1,n_2+1,nk;1)-\nonumber\\
&k(q_1+n)e(n_1,n_2+1,n,k-1;2)]
\label{52}
\end{eqnarray}

\section{Second quantization of representation operators}
\label{S6}

In conventional quantum theory the operators of 
physical quantities act in the Fock space of the system under 
consideration. Suppose, for example, that the system consists
of free superparticles and their antiparticles.
Since $(n_1n_2nk;i)$ is the full set of quantum numbers
characterizing the superparticles and their antiparticles, one
can define creation and annihilation operators for them.
Let
$$a(n_1n_2nk;i),\quad a(n_1n_2nk;i)^*, 
\quad b(n_1n_2nk;i)\quad b(n_1n_2nk;i)^*$$
be the operators having the meaning of annihilation operator 
for particle $i$ in the state with the quantum numbers $(n_1n_2nk)$,
creation operator 
for particle $i$ in the state with these quantum numbers,
annihilation operator 
for antiparticle $i$ in the state with these quantum numbers,
and creation operator 
for antiparticle $i$ in the state with these quantum numbers, respectively.
The basis of the Fock space consists of the vacuum vector $\Phi_0$ and
linear combinations of vectors obtained by acting on $\Phi_0$ by 
the creation operators for particles and antiparticles with all
possible quantum numbers.

In the standard approach, the necessity to have separate annihilation
and creation operators for particles and antiparticles, respectively,
is related to the fact that they are described by independent IRs of
the symmetry algebra. On the contrary, in the GFQT one IR describes
a particle and its antiparticle simultaneously. For this reason there
is no need to introduce $(b,b^*)$ operators since the set of all
possible operators $(a,a^*)$ describes annihilation and creation of
both particles and antiparticles. This question has been already
discussed in detail in Refs. \cite{lev2,lev3}.

The problem of second quantization of representation operators can
now be formulated as follows. Let $(A_1,A_2....A_n)$ be representation
generators describing IR of an algebra or superalgebra. One should
replaced them by operators acting in the Fock space such that the
commutation-anticommutation relations between the new operators are
the same. 

Let $A$ be a representation generator of the osp(1,4) algebra.
Its matrix elements are defined as
\begin{equation}
Ae(n_1n_2nk;i)=\sum_{n_1'n_2'n'k'i'}
A(n_1'n_2'n'k';i'|n_1n_2nk;i)e(n_1'n_2'n'k';i')
\label{53}
\end{equation}
If $B$ is the operator adjoint to $A$ then, as follows from 
the relation $(x,Ay)=(Bx,y)$, the matrix elements of these 
operators are related to each other as
\begin{eqnarray}
&B(n_1'n_2'n'k';i'|n_1n_2nk;i)=
[Norm(n_1n_2nk;i)/Norm(n_1'n_2'n'k';i')]\nonumber\\
&\overline{A(n_1n_2nk;i|n_1'n_2'n'k';i')}
\label{54}
\end{eqnarray}
This relation differs from the usual one since our basis
elements are not normalized to one.

We assume that the operators $a(n_1n_2nk;i)$ and $a(n_1n_2nk;i)^*$
satisfy either the anticommutation relation
\begin{equation}
\{a(n_1n_2nk;i),a(n_1'n_2'n'k';i')^*\}=Norm(n_1n_2nk;i) \delta_{n_1n_1'} 
\delta_{n_2n_2'} \delta_{nn'} \delta_{kk'} \delta_{ii'}
\label{55}
\end{equation}
or the commutation relation
\begin{equation}
[a(n_1n_2nk;i),a(n_1'n_2'n'k';i')^*]=Norm(n_1n_2nk;i) \delta_{n_1n_1'} 
\delta_{n_2n_2'} \delta_{nn'} \delta_{kk'} \delta_{ii'}
\label{56}
\end{equation}
In that case $\{a,a'\}=\{a^*,a^{'*}\}=0$ for any pair of
$a$ or $a^*$ operators in the case of Eq. (\ref{55}) and
$[a,a']=[a^*,a^{'*}]=0$ in the case of Eq. (\ref{56}).
Then the secondly quantized form of the operator $A$ is
\begin{eqnarray}
&A=\sum A(n_1'n_2'n'k';i'|n_1n_2nk;i)a(n_1'n_2'n'k';i')^*\nonumber\\
&a(n_1n_2nk;i)/Norm(n_1n_2nk;i)
\label{57}
\end{eqnarray}
where the sum is taken over all possible values of 
$(n_1'n_2'n'k'i'n_1n_2nki)$. The operator adjoint to $A$ in
the Fock space is obviously
\begin{eqnarray}
&A^*=\sum \overline{A(n_1'n_2'n'k';i'|n_1n_2nk;i)}
a(n_1n_2nk;i)^*\nonumber\\
&a(n_1'n_2'n'k';i')/Norm(n_1n_2nk;i)
\label{58}
\end{eqnarray}
At the same time, if $B$ is the operator adjoint to $A$ in
the space of IR then its secondly quantized form is
\begin{eqnarray}
&B=\sum B(n_1'n_2'n'k';i'|n_1n_2nk;i)a(n_1'n_2'n'k';i')^*\nonumber\\
&a(n_1n_2nk;i)/Norm(n_1n_2nk;i)
\label{59}
\end{eqnarray}
As follows from Eq. (\ref{54}), the operators given by
Eqs. (\ref{58}) and (\ref{59}), are equal to each other
and therefore the correspondence between the operators
in the space of IR and the Fock space is compatible with the
Hermitian conjugation in these spaces.

Suppose now that $A$ and $B$ are arbitrary
operators in the space of IR, and Eqs. (\ref{57}) and
(\ref{59}) represent their secondly quantized forms,
respectively. Then, if $A$ and $B$ satisfy some commutation
relation in the space of IR, the corresponding secondly 
quantized operators satisfy the same commutation relation
regardless whether the $(a,a^*)$ operators satisfy
Eq. (\ref{55}) or Eq. (\ref{56}). This means that for
representations of ordinary Lie algebras and their
modular analogs, the above quantization procedure by
itself does not impose any restrictions on the type of
statistics.

Let now $A$ and $B$ be two fermionic operators in IR
of the osp(1,4) superalgebra and Eqs. (\ref{57}) and
(\ref{59}) be their secondly quantized forms. Since 
the anticommutator of $A$ and $B$ is a linear combination
of generators of the so(2,3) algebra, the question
arises when the same relation is valid for the corresponding
secondly quantized operators. In other words, if $C$ is an
operator in the space of IR, $\{A,B\}=C$ in this space,
and the secondly quantized form of $C$ is
\begin{eqnarray}
&C=\sum C(n_1'n_2'n'k';i'|n_1n_2nk;i)a(n_1'n_2'n'k';i')^*\nonumber\\
&a(n_1n_2nk;i)/Norm(n_1n_2nk;i)
\label{60}
\end{eqnarray}
then the question arises when $\{A,B\}=C$ for the
secondly quantized operators $A$, $B$ and $C$ given by
Eqs. (\ref{57}), (\ref{59}) and (\ref{60}), respectively.

It is easy to see that this relation cannot be satisfied
if the $(a,a^*)$ operators satisfy either Eq. (\ref{55})
or Eq. (\ref{56}) for all the values of $i$ and $i'$.
This means that supersymmetry indeed combines 
particles with different type of statistics 
into one supermultiplet. For a short time, we renumerate 
the particles as follows: particles 0 and
2 will refer to bosons while particles 1 and 3 - to
fermions. Assuming that bosonic and fermionic operators
commute with each other, we can write the relations
between all the $(a,a^*)$ operators as follows
\begin{eqnarray}
&a(n_1n_2nk;i)a(n_1'n_2'n'k';i')^*=
Norm(n_1n_2nk;i) \delta_{n_1n_1'} 
\delta_{n_2n_2'} \delta_{nn'} \delta_{kk'} \delta_{ii'}\nonumber\\
&+(-1)^{ii'}a(n_1'n_2'n'k';i')^*a(n_1n_2nk;i)\nonumber\\
&a(n_1n_2nk;i)^*a(n_1'n_2'n'k';i')^*=
(-1)^{ii'}a(n_1'n_2'n'k';i')^*a(n_1n_2nk;i)^*\nonumber\\
&a(n_1n_2nk;i)a(n_1'n_2'n'k';i')=
(-1)^{ii'}a(n_1'n_2'n'k';i')a(n_1n_2nk;i)
\label{61}
\end{eqnarray} 
Then one can directly verify the validity of the following 

{\it Statement 2}:
if $A$, $B$ and $C$ are operators in the 
representation space of the osp(1,4) algebra such that
$\{A,B\}=C$, then the secondly quantized forms of these
operators given by Eqs. (\ref{58}-\ref{60}) satisfy the
same relation in the Fock space if and only if the 
operators $A$ and $B$ have nonzero matrix elements only for 
transitions $boson\rightarrow fermion$ and 
$fermion\rightarrow boson$.

Returning to the initial way of enumerating the particles
in the supermultiplet (see Sect. \ref{S4}), we conclude
that, as a consequence of Eqs. (\ref{45}-\ref{52})): if
particle 0 is a boson then particles 1 and 2 are fermions
and particle 3 is a boson while if particle 0 is a
fermion then particles 1 and 2 are bosons and particle 3
is a fermion. As follows from these expressions and the
discussed rule of constructing secondly quantized
operators, the secondly quantized forms of the operators
$d_1^*$ and $d_2^*$ can be written as follows:
\begin{eqnarray}
d_1^*=d_1^{*(0)}+d_1^{*(1)}+d_1^{*(2)}+d_1^{*(3)}\quad
d_2^*=d_2^{*(0)}+d_2^{*(1)}+d_2^{*(2)}+d_2^{*(3)}
\label{62}
\end{eqnarray}
where
\begin{eqnarray}
&d_1^{*(0)}=\sum \{\frac{s+1-k}{s+1}a(n_1n_2nk;1)^*-
ka(n_1n_2,n,k-1;2)^*+\nonumber\\
&\frac{n(q_2+n-2)}{(q_1+n-k-1)
(q_1+n-k-2)(s+1)}a(n_1+1,n_2,n-1,k+1;1)^*+\nonumber\\
&\frac{n(q_1+n-1)}{(q_1+n-k-1)(q_1+n-k-2)}
a(n_1+1,n_2,n-1,k;2)^*\}\nonumber\\
&a(n_1n_2nk;0)/Norm(n_1n_2nk;0)
\label{63}
\end{eqnarray}
\begin{eqnarray}
&d_1^{*(1)}=\sum\{\frac{k(q_2+k-2)}{m-2}
a(n_1n_2,n+1,k-1;0)^*-\nonumber\\
&ka(n_1n_2n,k-1;3)^*+\nonumber\\
&\frac{(m+n-2)(q_1+n)(q_1-k-1)}{(m-2)(q_1+n-k)(q_1+n-k-1)}
a(n_1+1,n_2nk;0)^*+\nonumber\\
&\frac{n(q_1+n)}{(q_1+n-k)(q_1+n-k-1)}
a(n_1+1,n_2,n-1,k;3)^*\}\nonumber\\
&a(n_1n_2nk;1)/Norm(n_1n_2nk;1)
\label{64}
\end{eqnarray}
\begin{eqnarray}
&d_1^{*(2)}=\sum\{\frac{(s-k)(q_2+k+1)}{(s+1)(m-2)}
a(n_1n_2,n+1,k;0)^*-\frac{s-k}{s+1}a(n_1n_2nk;3)^*-\nonumber\\
&\frac{(q_2+n-1)(m+n-2)(q_1-k-2)}{(q_1+n-k-1)(q_1+n-k-2)((s+1)(m-2)}
a(n_1+1,n_2,n,k+1;0)^*-\nonumber\\
&\frac{n(q_2+n-1))}{(q_1+n-k-1)(q_1+n-k-2)(s+1)}
a(n_1+1,n_2,n-1,k+1;3)\}\nonumber\\
&a(n_1n_2nk;2)/Norm(n_1n_2nk;2)
\label{65}
\end{eqnarray}
\begin{eqnarray}
&d_1^{*(3)}=\sum\{\frac{(q_2+k-1)(s+1-k)}{(m-2)(s+1)}
a(n_1n_2,n+1,k;1)^*-\nonumber\\
&\frac{k(q_2+k-1)}{m-2}
a(n_1n_2,n+1,k-1;2)^*-\nonumber\\
&\frac{(m+n-1)(q_1-k-1)(q_2+n-1)}{(m-2)(q_1+n-k)(q_1+n-k-1)(s+1)}
a(n_1+1,n_2,n,k+1;1)^*-\nonumber\\
&\frac{(m+n-1)(q_1-k-1)(q_1+n)}{(m-2)(q_1+n-k)(q_1+n-k-1)}
a(n_1+1,n_2,n,k;2)^*\}\nonumber\\
&a(n_1n_2nk;3)/Norm(n_1n_2nk;3)
\label{66}
\end{eqnarray}
\begin{eqnarray}
&d_2^{*(0)}=\sum\{a(n_1n_2nk;2)^*+\frac{1}{s+1}
a(n_1n_2n,k+1;1)^*+\nonumber\\
&\frac{n(s+1-k)(q_2+n-2)}{(s+1)(q_2+n+k-1)(q_2+n+k-2)}
a(n_1,n_2+1,n-1,k;1)^*-\nonumber\\
&\frac{kn(q_1+n-1)}{(q_2+n+k-1)(q_2+n+k-2)}
a(n_1,n_2+1,n-1,k-1;2)^*\}\nonumber\\
&a(n_1n_2nk;0)/Norm(n_1n_2nk;0)
\label{67}
\end{eqnarray}
\begin{eqnarray}
&d_2^{*(1)}=\sum\{\frac{q_1-k-1}{m-2}e(n_1n_2,n+1,k;0)+
a(n_1n_2nk;3)^*+\nonumber\\
&\frac{k(q_1+n)(m+n-2)(q_2+k-2)}{(m-2)(q_2+n+k-1)(q_2+n+k-2)}
a(n_1,n_2+1,n,k-1;0)^*-\nonumber\\
&\frac{kn(q_1+n)}{(q_2+n+k-1)(q_2+n+k-2)}
a(n_1,n_2+1,n-1,k-1;3)^*\}\nonumber\\
&a(n_1n_2nk;1)/Norm(n_1n_2nk;1)
\label{68}
\end{eqnarray}
\begin{eqnarray}
&d_2^{*(2)}=\sum\{-\frac{q_1-k-2}{(s+1)(m-2)}
a(n_1,n_2,n+1,k+1;0)^*-\nonumber\\
&\frac{1}{s+1}a(n_1n_2n,k+1;3)^*+\nonumber\\
&\frac{(s-k)(m+n-2)(q_2+n-1)(q_2+k-1)}{(s+1)(m-2)(q_2+n+k)(q_2+n+k-1)}
a(n_1,n_2+1,nk;0)^*-\nonumber\\
&\frac{n(s-k)(q_2+n-1)}{(s+1)(q_2+n+k)(q_2+n+k-1)}
a(n_1,n_2+1,n-1,k;3)^*\}\nonumber\\
&a(n_1n_2nk;2)/Norm(n_1n_2nk;2)
\label{69}
\end{eqnarray}
\begin{eqnarray}
&d_2^{*(3)}=\sum\{-\frac{q_1-k-1}{m-2}
a(n_1,n_2,n+1,k;2)^*-\nonumber\\
&\frac{q_1-k-1}{(m-2)(s+1)}
a(n_1n_2,n+1,k+1;1)^*+\nonumber\\
&\frac{(m+n-1)(q_2+k-1)}{(m-2)(q_2+n+k-1)(q_2+n+k)}
[\frac{(q_2+n-1)(s+1-k)}{s+1}a(n_1,n_2+1,nk;1)^*-\nonumber\\
&k(q_1+n)a(n_1,n_2+1,n,k-1;2)]\}\nonumber\\
&a(n_1n_2nk;3)/Norm(n_1n_2nk;3)
\label{70}
\end{eqnarray}
where the sum in each expression is taken over all
possible values of the quantities $(n_1n_2nk)$.
Using these equations one can easily write down
(if necessary) the secondly quantized forms of the
operators $d_1$ and $d_2$.

The reader can notice that the results of this and 
preceding sections are obtained without explicitly
using the fact that we consider the theory over a
Galois field. The analogous results take place in the
standard theory as well. The difference between the two
approaches is that in the standard one the quantities
$(n_1n_2n)$ take the infinite number of values while
in the GFQT - only the finite ones. 
  
\section{AB symmetry of fermionic operators}
\label{S7}

One can reformulate the result (\ref{21}) as follows. 
For each $i$, $n$ and $k$, the quantities $n_1$ and $n_2$ 
take the values
\begin{eqnarray}
&n_1=0,1,...N_1(n,k;i)\quad n_2=0,1,...N_2(n,k;i)\nonumber\\
&N_1(n,k;i)=p-q_{1i}-n+k\quad N_2(n,k;i)=p-q_{2i}-n-k
\label{71}
\end{eqnarray}
Then, as follows from the results of Ref. \cite{lev3},
the secondly quantized so(2,3) generators for particle $i$ 
are invariant under the AB transformation 
\begin{eqnarray}
&a(n_1n_2nk;i)^*\rightarrow (-1)^{n_1+n_2+n}\alpha_i\nonumber\\ 
&a(N_1(n,k;i)-n_1,N_2(n,k;i)-n_2,nk;i)\nonumber\\
&F(N_1(n,k;i)-n_1,N_2(n,k;i)-n_2,nk;i)^{-1}\nonumber\\ 
&a(n_1n_2nk;i)\rightarrow (-1)^{n_1+n_2+n} 
{\bar\alpha}_i\nonumber\\
&a(N_1(n,k;i)-n_1,N_2(n,k;i)-n_2,nk;i)^*\nonumber\\
&F(N_1(n,k;i)-n_1,N_2(n,k;i)-n_2nk;i)^{-1}          
\label{72}
\end{eqnarray}
if the constant $\alpha_i$, which can be called the 
AB parity of particle $i$, satisfies the condition
\begin{equation}
\alpha_i{\bar\alpha}_i=\pm (-1)^{s_i}
\label{73}
\end{equation} 
with the plus sign for fermions and the minus sign for
bosons. 

If the spin-statistics theorem is applicable
to the GFQT (i.e. we have a property that
the fermions have an odd spin in our units and the
bosons - the even one) then Eq. (\ref{73}) becomes 
\begin{equation}
\alpha_i{\bar\alpha}_i=-1
\label{74}
\end{equation}
for both, fermions and bosons. Such a relation is
impossible in the standard theory but is possible
if $\alpha_i\in F_{p^2}$. Indeed, we can use 
the fact that any Galois field is 
cyclic with respect to multiplication \cite{VDW}. 
Let $r$ be a primitive root of $F_{p^2}$. This means that any
element of $F_{p^2}$ can be represented as a power of $r$. 
As mentioned in Sect. \ref{S2}, $F_{p^2}$ has only one
nontrivial automorphism which is defined as 
$\alpha\rightarrow {\bar \alpha}=\alpha^p$. 
Therefore if $\alpha =r^k$ then $\alpha{\bar \alpha}=
r^{(p+1)k}$. On the other hand, since $r^{(p^2-1)}=1$, we
conclude that $r^{(p^2-1)/2}=-1$. Therefore there exists at
least a solution with $k=(p-1)/2$. 

Now the following remarks are in order. We assume that 
the AB symmetry is the fundamental symmetry in the 
GFQT. This means in particular that in
the supersymmetric version of the GFQT the generators
obtained after the transformation (\ref{72}), satisfy
the same relations as the
original generators. This property is obviously
satisfied if all the generators, including the
fermionic ones, are invariant under this 
transformation. However, the relations (\ref{30}) 
also remain unchanged if the AB transformation
transforms $(d_1,d_2,d_1^*,d_2^*)$ into
$(-d_1,-d_2,-d_1^*,-d_2^*)$. In other words, while
the generators of the so(2,3) algebra should
necessarily have the AB parity equal to 1, the
fermionic generators can have the AB parity equal
to either 1 or -1. At this stage it is not clear 
which of the possibilities should be accepted. For
this reason we investigate for definiteness the
conditions when the fermionic generators are
invariant under the AB transformation, i.e. their
AB parity is equal to 1. It will be clear that the
results can be easily reformulated if the AB parity
of the fermionic generators should be equal to -1. 

By using the famous Wilson theorem in number
theory that $(p-1)!=-1\,\, (mod\,\, p)$ in $F_p$
(see e.g. Ref. \cite{VDW}) and Eqs. (\ref{40})
and (\ref{71}), one can easily show that
\begin{equation}
F(N_1(n,k;i)-n_1,N_2(n,k;i)-n_2,nk;i)F(n_1,n_2nk;i)=(-1)^{s_i}
\label{75}
\end{equation} 
This fact is very important in investigating the
AB symmetry. 

Consider the first term in Eq. (\ref{63})
\begin{eqnarray}
&d_{1(1)}^{*(0)}=\sum \frac{s+1-k}{s+1}a(n_1n_2nk;1)^*
a(n_1n_2nk;0)/Norm(n_1n_2nk;0)
\label{76}
\end{eqnarray}
As follows from the results of Sect. \ref{S3}, the
quantities $(n_1n_2nk)$ in this sum are in the range
\begin{equation}
n_1\in [0,N_1(n,k;1)]\,\, n_2\in [0,N_2(n,k;0)] \,\,
 n\in [0,p-3-m]\,\, k\in [0,s]
\label{77}
\end{equation}
since $q_{11}=q_1+1$, $q_{21}=q_2$, $s_1=s+1$ and 
$m_1=m+1$.

Now we apply the transformation (\ref{72}). Assuming that the
bosonic and fermionic operators commute with each other
and using Eqs. (\ref{40}) and (\ref{71}), we obtain that 
the transformed form of this expression is
\begin{eqnarray}
&(d_{1(1)}^{*(0)})^{AB}=\alpha_1{\bar\alpha}_0
\sum [(s+1-k)/(s+1)]\nonumber\\
&a(N_1(n,k;0)-n_1,N_2(n,k;0)-n_2,nk;0)^*\nonumber\\
&a(N_1(n,k;1)-n_1,N_2(n,k;1)-n_2,nk;1)\nonumber\\
&[F(N_1(n,k;1)-n_1,N_2(n,k;1)-n_2,nk;1)\nonumber\\
&F(N_1(n,k;0)-n_1,N_2(n,k;0)-n_2,nk;0)\nonumber\\
&F(n_1n_2nk;0)G(nk;0)]^{-1}
\label{78}
\end{eqnarray}
By using Eq. (\ref{75}),
we can rewrite this expression as 
\begin{eqnarray}
&(d_{1(1)}^{*(0)})^{AB}=(-1)^s\alpha_1{\bar\alpha}_0
\sum [(s+1-k)/(s+1)]\nonumber\\
&a(N_1(n,k;0)-n_1,N_2(n,k;0)-n_2,nk;0)^*\nonumber\\
&a(N_1(n,k;1)-n_1,N_2(n,k;1)-n_2,nk;1)\nonumber\\
&[F(N_1(n,k;1)-n_1,N_2(n,k;1)-n_2,nk;1)G(nk;0)]^{-1}
\label{79}
\end{eqnarray}

As follows from Eq. (\ref{40})
\begin{equation}
G(nk;0)=G(nk;1)\frac{(s+1-k)(m-2)(q_1+n-k)(q_1+n-k-1)}
{(s+1)(m+n-2)(q_1+n)(q_1-k-1)}
\label{80}
\end{equation}
By using this relation, changing the summation
variables in Eq. (\ref{79}) as 
$n_1\rightarrow N_1(n,k;1)-n_1$,
$n_2\rightarrow N_2(n,k;1)-n_2$ and using the
relations $N_1(n,k;0)=N_1(n,k;1)+1$ and
$N_2(n,k;0)=N_2(n,k;1)$ (see Eq. (\ref{71})) we 
arrive at the following conclusion:
the AB transformation of the first term in Eq. (\ref{63})
is equal to the third term in Eq. (\ref{64}) if
$(-1)^s\alpha_1{\bar\alpha}_0=1$.

The investigation of the other terms in Eqs. 
(\ref{63}-\ref{70}) can be carried out analogously, and
the result is as follows. The fermionic generators are
invariant under the AB transformation if
\begin{eqnarray}
\alpha_3=-\alpha_1\,\, \alpha_2=\alpha_1\,\, 
(-1)^s\alpha_1{\bar\alpha}_0=(-1)^s\alpha_0{\bar\alpha}_1=1
\label{81}
\end{eqnarray}

Consider the last relation in Eq. (\ref{81}). Since the
automorphism in $F_{p^2}$ is defined as 
$\alpha \rightarrow \alpha^p$ (see Sect. \ref{S2}), we
conclude that if $\alpha_1=\alpha_0\kappa$ then 
$\kappa^{p-1}$ =1. This relation shows that 
$\kappa$ is an element of $F_p$. Assuming that the
spin-statistics relation is satisfied (see Eq. (\ref{74})),
our final conclusion can be formulated in the form
of

{\it Statement 3}: The fermionic generators in the
massive supermultiplet are invariant
under the AB transformation if and only if the AB
parities of the particles in the supermultiplet satisfy 
the relations
\begin{eqnarray}
\alpha_3=-\alpha_0\quad \alpha_2=\alpha_1\quad 
\alpha_1=\alpha_0\kappa\quad \kappa =(-1)^{s+1}
\label{82}
\end{eqnarray}
The fermionic generators change their sign under the
AB transformation if the last relation is replaced by
$\kappa=(-1)^s$.

The investigation of the collapsed supermultiplet,
massless supermultiplet and Dirac supermultiplet
(see Sect. \ref{S4}) can be carried out analogously
and simpler. In the collapsed supermultiplet,
particle 2 is absent and the remaining relations
between the AB parities are the same as in Eq. (\ref{82}).
In the massles and Dirac supermultiplets, particles
2 and 3 are absent, and the relation between the
AB parities of particles 0 and 1 is the same as in
Eq. (\ref{82}).

\section{Space inversion in GFQT}
\label{S8} 

In terms of representation generators of the so(2,3)
algebra, the space inversion is defined as a
transformation $M_{ab}\rightarrow M_{ab}^P$ such that
\begin{equation}
M_{ik}^P=M_{ik}\quad M_{i0}^P=-M_{i0} \quad 
M_{i5}^P=-M_{i5}\quad M_{05}^P=M_{05}\quad (i,k=1,2,3) 
\label{83}
\end{equation}
Our first task is to find a transformation of the
$(a,a^*)$ operators resulting in Eq. (\ref{83}).
It is easy to see that if the transformation is defined as 
\begin{eqnarray}
&a(n_1n_2nk;i)^*\rightarrow (-1)^{n_1+n_2+n}{\bar\eta}_i 
a(n_1n_2nk;i)^*\nonumber\\
&a(n_1n_2nk;i)\rightarrow (-1)^{n_1+n_2+n}\eta_i 
a(n_1n_2nk;i)
\label{84}
\end{eqnarray}
where the parity $\eta_i$ of particle $i$ is such that
$\eta_i{\bar\eta}_i=1$ then the transformed generators
for particle $i$ indeed satisfy Eq. (\ref{83}). Indeed,
as follows from the construction described in Sect. 
\ref{S3} (see also Ref. \cite{lev3}) for details),
the operators $(a_j',a_j",b',b")$ $(j=1,2)$ have only
such nonzero matrix elements for which the values of 
$n_1+n_2+n$ in the initial and final states differ by
$\pm 1$. Therefore these operators change their sign
under transformation (\ref{84}). At the same time, the
operators $L_{\pm},h_j$ have nonzero matrix elements
only for transitions where the values of $n_1+n_2+n$
in the initial and final states are the same. 
Then the result (\ref{83}) follows from Eq. (12).

A well known result in the conventional theory is that
fermions have imaginary parity (see e.g. Ref. 
\cite{Wein-super}). We will show soon that in 
the GFQT this result is a consequence of the AB symmetry. 
However, we will first discuss a well known mathematical
problem of how the field $F_p$ can be extended to 
$F_{p^2}$ (see e.g. Ref. \cite{VDW}).

By analogy with the field of complex numbers, we can 
try to define $F_{p^2}$ as a set 
of $p^2$ elements $a+bi$ where $a,b\in F_p$ and $i$ is a 
formal element such that $i^2=1$. The question arises whether 
so defined $F_{p^2}$ is a field, i.e. we can define all the 
four operations excepting division by zero.
The definition of addition, subtraction and multiplication 
in so defined $F_{p^2}$ is obvious and, by analogy with the 
field of complex numbers, we can
try to define division as $1/(a+bi)\,=a/(a^2+b^2)\,-ib/(a^2+b^2)$.
This definition can be meaningful only if $a^2+b^2\neq 0$ in $F_p$
for any $a,b\in F_p$ i.e. $a^2+b^2$ is not divisible by $p$.
Therefore the definition is meaningful only if $p$ {\it cannot}
be represented as a sum of two squares and is meaningless otherwise.
Since the prime number $p$ in question is necessarily odd, we
have two possibilities: the value of $p\,(mod \,4)$ is either 1
or 3. The well known result of the number theory (see e.g. the 
textbooks \cite{VDW}) is that a prime number $p$ can be 
represented as a sum of two squares only in the former case
and cannot in the latter one. Therefore the above construction of
the field $F_{p^2}$ is correct only if $p\,(mod \,4)\,=\,3$.
In that case it is easy to verify that if $z=a+ib$ then the
automorphism $z\rightarrow {\bar z}=z^p$ is the usual complex
conjugation.

In the general case we can extend $F_p$ to $F_{p^2}$ as 
follows. Let the equation
$\kappa^2=-a_0$ $(a_0\in F_p)$ have no solutions in $F_p$.
Then $F_{p^2}$ can be formally described as a set of
elements $a+b\kappa$, where $a,b\in F_p$ and $\kappa$ 
satisfies the condition $\kappa^2=-a_0$. The actions in
$F_{p^2}$ are defined in the natural way. The condition
that the equation $\kappa^2=-a_0$ has no solutions in 
$F_p$ is important in order to ensure that any nonzero 
element from $F_{p^2}$ has an inverse.
Indeed, the definition 
$(a+b\kappa)^{-1}=(a-b\kappa)/(a^2+b^2a_0)$ is correct 
since the denominator can be equal to zero only if both,
$a$ and $b$ are equal to zero. 
  
If $p=3\, (mod \, 4)$ then a possible choice of $a_0$ is
$a_0=1$ and then one can use the usual notation $i$ for 
$\kappa$. If $p=1\, (mod \, 4)$ then the choice $a_0=1$
is impossible but other possible choices exist and
$\overline{a+b\kappa}=a-b\kappa$ (see Ref.
\cite{lev2} for details). In the both cases it is 
possible to prove the correspondence between the GFQT and 
the standard approach (see Ref. \cite{lev2}). In other
words, at present one {\it cannot} unambigously conclude that
the realistic value of $p$ is only such that 
$p=3\, (mod \, 4)$.

We now return to the problem of fermion parity. Apply
transformation (\ref{84}) to the both parts
of the first (or second) expression in Eq. (\ref{72}).
Then, as follows from Eq. (\ref{71}), 
${\bar\eta}_i=(-1)^{s_i}\eta_i$. Therefore, in the both
cases, $p=1\, (mod \, 4)$  and $p=3\, (mod \, 4)$, the
parity of fermions is "imaginary" and the parity of bosons
is "real". However, the relation $\eta_i{\bar\eta}_i=1$
can be satisfied only in the second case. 

Consider now when the transformation defined by Eq. (\ref{84})
is compatible with supersymmetry. Suppose that, as a result
of this transformation, 
\begin{equation}
d_j\rightarrow \eta d_j\quad 
d_j^*\rightarrow {\bar\eta} d_j^* \quad(j=1,2)
\label{85}
\end{equation}
Then, as follows from Eq. (\ref{31}), the operators
$(a_j',a_j",b',b")$ will indeed change their sign 
if $\eta^2=-1$, and the operators $L_{\pm},h_j$ will remain
unchanged if $\eta{\bar\eta}=1$. Therefore, in the standard
theory the quantity $\eta$ should be imaginary, and the same
is valid in the GFQT if $p=3\, (mod \, 4)$ (see the above
discussion). Now, as follows from Eqs. (\ref{63}-\ref{70}) 
and (\ref{84}), the property (\ref{85}) can be satisfied
if and only if $\eta_2=\eta_1$, $\eta_3=-\eta_0$ 
(in the standard theory these properties are well known
 \cite{Wein-super}) and 
${\bar\eta}={\bar\eta}_1\eta_0=-{\bar\eta}_0\eta_1$. The last 
property is again satisfied if the parity of fermions is
"imaginary" and the parity of bosons is "real", in full
analogy with the standard theory.

\section{X inversion}
\label{S9}

The relations (\ref{30}) defining a representation of the
osp(1,4) algebra are obviously invariant under the
transformation $d_1\leftrightarrow d_2$,
$d_1^*\leftrightarrow d_2^*$. 
As follows from Eq. (\ref{31}), in this
case $a_1'\leftrightarrow a_2'$, 
$a_1"\leftrightarrow a_2"$, $h_1\leftrightarrow h_2$
and the operators $(L_{\pm},b',b")$ remain unchanged.
Therefore, as follows from Eq. (\ref{12}), if 
$a,b=0,2,3,5$ then $M^{1a}\rightarrow -M^{1a}$ and
$M^{ab}\rightarrow M^{ab}$. One can therefore 
treat the above transformation as that corresponding
to the inversion of the $x$ axis. Since the
conventional space inversion can be obtained by
successively applying the inversions of the $x$,
$y$ and $z$ axis, it is reasonable to think that
X inversion is more general than the space one. 

We will show that in terms of the $(a,a^*)$ operators
the X transformation can be defined as
\begin{eqnarray}
&a(n_1n_2nk;i)\rightarrow \beta_i \frac{k!}{(s_i-k)!}
a(n_2n_1n,s_i-k;i)\nonumber\\
&a(n_1n_2nk;i)^*\rightarrow \beta_i \frac{k!}{(s_i-k)!}
a(n_2n_1n,s_i-k;i)^*
\label{86}
\end{eqnarray} 
where $\beta_i$ is the X parity. The fact that $\beta_i$
is "real" easily follows from the AB symmetry (see Eq.
(\ref{72})). Note that, as follows from Eq. (\ref{71})
\begin{equation}
N_1(n,s_i-k;i)=N_2(n,k;i)\quad N_2(n,s_i-k;i)=N_1(n,k;i)
\label{87}
\end{equation} 
As follows from Eq. (\ref{40})
\begin{eqnarray}
&G(n,s_i-k;i)=[\frac{k!}{(s_i-k)!}]^2G(nk;i)\nonumber\\
&F(n_1n_2nk;i)=F(n_2n_1n,s_i-k;i) 
\label{88}
\end{eqnarray}
For this reason the anticommutation or commutation
relations (\ref{55}) or (\ref{56}) are invariant under
the transformation (\ref{86}) if $(\beta_i)^2=1$.

Consider again the first term in Eq. (\ref{63}), which,
as noted above, can be written in the form of Eq. 
(\ref{76}). After applying the transformation (\ref{86}),
this expression becomes
\begin{eqnarray}
&(d_{1(1)}^{*(0)})^X=\beta_0\beta_1
\sum \frac{1}{s+1}[\frac{k!}{(s-k)!}]^2
a(n_2n_1n,s+1-k;1)^*\nonumber\\
&a(n_2n_1n,s-k;0)[F(n_1n_2nk;0)G(nk;0)]^{-1}
\label{89}
\end{eqnarray}
since $s_1=s+1$. By using Eq. (\ref{87}), (\ref{88})
and changing the summation variables as 
$n_1\leftrightarrow n_2$, $k\rightarrow (s-k)$, it is
easy to demonstrate that this expression becomes the
first term in Eq. (\ref{67}) if $\beta_1=\beta_0$.
One can consider the other terms analogously and come to

{\it Statement 4:} The transformation (\ref{86}) is
a symmetry if
\begin{equation}
\beta_1=-\beta_2=-\beta_3=\beta_0
\label{90}
\end{equation}
We see that the X parities of particles 2 and 3 are
opposite to those for particles 0 and 1.

\section{Conclusion}
\label{S10}

The main result of the present paper, proved in Sect.
\ref{S7}, is that the AB symmetry (discussed in 
detail in Refs \cite{lev2,lev3}) is compatible with
supersymmetry. Let us discuss why this result is very
important (in our opinion). First of all we briefly
describe the meaning of the AB symmetry and explain 
why it has no analog in the standard theory.

In quantum theory based on a Galois field (GFQT),
the field of complex numbers is replaced by a Galois
field $F_{p^2}$ containing $p^2$ elements. In 
particular, the ring of integers $Z$ is replaced by
a simplest Galois field $F_p=Z/Zp$ --- the residue
field modulo $p$. While the elements from $Z$ are
in the range $-\infty,...-2,-1,0,1,2,...\infty$,
one can treat $F_p$ as a set of elements in the range
$-A,...-2,-1,0,1,2,...A$ where $A=(p-1)/2$. One might
think that if $p$ is very big, there should not be
a big difference between the theories based on $Z$ 
and $F_p$. However, since arithmetic of Galois fields
differs from the standard one, it is reasonable to
expect that the GFQT has its own specific features.
As already noted, in the GFQT a particle and its
antiparticle belong to the same IR of the symmetry
algebra, in contrast with the standard theory where
they belong to independent IRs. The AB symmetry 
relates particles and antiparticles within their
common IR, and that's why it has no analog in the
standard theory.

The existence of the AB symmetry imposes considerable
restrictions on the structure of the GFQT \cite{lev3}.
Since any osp(1,4) supermultiplet contains 2, 3 or 4
particles described by modular IRs of the so(2,3)
algebra, the conclusions of Ref. \cite{lev3} remain
unchanged in the presence of supersymmetry. The fact
that the AB supersymmetry passes the supersymmetry test
is a strong argument in favor of our assumption that
it plays a fundamental role. If this is accepted 
then in the presence
of supersymmetry it is possible to impose additional
restrictions on the structure of interactions. In
particular, the AB parities of particles in a
supermultiplet are not independent (see Sect. \ref{S7}).

In Sects. \ref{S8} and \ref{S9} we have considered space
inversion and X inversion in the GFQT. In particular,
we have reproduced the well known fact that the parities
of bosons are real and the parities of fermions are 
imaginary \cite{Wein-super}. This result is interesting
for two reasons. First, in the GFQT it is a direct
consequence of the AB symmetry. In addition (see the
discussion in Sect. \ref{S8}), it gives an indication
that $p=3\,\, (mod\,\, 4)$ rather than 
$p=1\,\, (mod\,\, 4)$ since only in the former case
$F_{p^2}$ contains an element with the same properties as
the imaginary unity $i$. If one accepts that the GFQT is
fundamental, the result can be reinterpreted in this
way: quantum theory involves complex numbers since 
(for some reasons) $p=3\,\, (mod\,\, 4)$.


\begin{thebibliography}{99}
\bibitem{lev1} F.M. Lev, Yad. Fiz. {\bf 48}, 903 (1988); 
J. Math. Phys. {\bf 30}, 1985 (1989); J. Math. Phys. {\bf 34}, 
490 (1993).
\bibitem{lev2} F.M. Lev, hep-th/0206078.
\bibitem{lev3} F.M. Lev, hep-th/0207192, hep-th/0209001.
\bibitem{Pauli} W. Pauli, Phys. Rev. {\bf 58}, 116 (1940).
\bibitem{Heidenreich} W. Heidenreich, Phys. Lett. 
{\bf B110}, 461 (1982).
\bibitem{VDW} B.L. Van der Waerden, "Algebra I", 
(Springer-Verlag, Berlin Heidelberg New York, 1967);
K. Ireland and  M. Rosen, "A Classical 
Introduction to Modern Number Theory", Graduate Texts in 
Mathematics-87, (New  York  - Heidelberg - Berlin: Springer, 1987); 
H. Davenport, "The Higher Arithmetic", 
(Cambridge University Press, 1999).
\bibitem{FrPa} E.M. Friedlander and B.J. Parshall, Bull. Am.
Math. Soc. {\bf 17}, 129 (1987).
\bibitem{Zass} H. Zassenhaus, Proc. Glasgow Math. Assoc. 
{\bf 2}, 1 (1954).
\bibitem{Fronsdal} C. Fronsdal, Rev. Mod. Phys. 
{\bf 37}, 221 (1965).
\bibitem{Evans} N.T. Evans, J. Math. Phys. 
{\bf 8}, 170 (1967). 
\bibitem{Braden} B. Braden, Bull. Amer. Math. Soc. 
{\bf 73}, 482 (1967);
Thesis, Univ. of Oregon, Eugene, OR (1966).
\bibitem{IW} E. Inonu and E.P. Wigner, Nuovo Cimento, 
{\bf IX}, 705 (1952).
\bibitem{DiracS} P.A.M. Dirac, J. Math. Phys. 
{\bf 4}, 901 (1963).
\bibitem{Wein-super} S. Weinberg, "The Quantum 
Theory of Fields", Volume III Supersymmetry, (Cambridge 
University Press, Cambridge, 2000).
\end{thebibliography}
\end{document}